\documentclass[twocolumn,prd,superscriptaddress,showpacs,amsmath,amssymb]{revtex4-1}

\usepackage{graphicx}
\usepackage{dcolumn}
\usepackage{bm}
\usepackage{multirow}
\usepackage{enumitem}

\begin{document}

\title{Liquid Scintillator Response to  Proton Recoils in the 10-100 keV Range}

\author{C. \!Awe}
\affiliation{Department of Physics, and Triangle Universities Nuclear Laboratory, Duke University, Durham, NC 27708, USA.}

\author{P.S. \!Barbeau}
\affiliation{Department of Physics, and Triangle Universities Nuclear Laboratory, Duke University, Durham, NC 27708, USA.}

\author{J.I. \!Collar}
\thanks{Corresponding author: collar@uchicago.edu}
\affiliation{Enrico Fermi Institute, Kavli Institute for Cosmological Physics, and Department of Physics, University of Chicago, Chicago, IL 60637, USA}

\author{S. \!Hedges}
\affiliation{Department of Physics, and Triangle Universities Nuclear Laboratory, Duke University, Durham, NC 27708, USA.}

\author{L. \!Li}
\affiliation{Department of Physics, and Triangle Universities Nuclear Laboratory, Duke University, Durham, NC 27708, USA.}

\date{\today}

\begin{abstract}
We study the response of EJ-301 liquid scintillator  to monochromatic 244.6 $\pm$ 8.4 keV neutrons, targeting the 10-100 keV proton recoil energy interval. Limited experimental information exists for proton light yield in this range, for this or any other organic scintillator. Our results confirm the adequacy of a modified Birks' model, common to all organic scintillator formulations, predicting a marked increase in quenching factor as proton energy approaches the few keV regime. The relevance of this behavior within the context of searches for low-mass particle dark matter is mentioned.  
\end{abstract}



\maketitle

\section{Motivation}

Thirty years of direct searches for dark matter particles, heavily focused on hypothetical Weakly Interacting Massive Particles (WIMPs) \cite{goodman,wasserman,wimps1,wimps2}, have returned no unambiguous positive detection. As a reaction to this, a number of new initiatives seek to expand the reach of these efforts. In particular, recent phenomenological work has concentrated on particle models involving low-mass candidates \cite{models}, incapable or limited in their ability to produce signals above the energy threshold of existing WIMP detectors. 

For dark matter (DM) particle masses of order 1 GeV/c$^{2}$, the match in mass of the projectile and a proton target in hydrogenated organic scintillator results in an increase in expected interaction rate, and in the maximum proton recoil energy that can be imparted during a DM-nucleus collision. This is advantageous in the context of searches for low-mass Strongly Interacting Massive Particles (SIMPs), a DM possibility that has recently attracted  renewed interest \cite{simp1,simp2,simp3,mypaper}. Furthermore, there are indications that a favorably large ($>$20\%) quenching factor (QF) might be in place for the proton recoil energies of order 1 keV that would be involved. This QF expresses the ratio between the light yield observed from a nuclear recoil, and from an electron recoil of the same energy. A value of the QF this large would ensure the generation of detectable levels of scintillation light from still-unexplored regions of low-mass SIMP parameter space \cite{mypaper}. For a variety of organic scintillators, this QF has nevertheless been repeteadly observed to monotonically decrease with decreasing proton energy, reaching a minimum of  $\sim$5\% around 100 keV. 

The change in this trend, towards the mentioned QF $>$ 20\% in the 0.1-1 keV proton energy range, has  been observed by  one group of experimenters only, during neutron scattering calibrations  employing NE-110 as the target \cite{ahlen1,ahlen2}. This organic plastic scintillator has been used for magnetic monopole searches \cite{mm7,mm8}. In these calibrations, the expected average scintillation signal was well-below one single photoelectron (SPE). As a result, less-than-straightforward statistical methods were involved in the determination of the quenching factor.  

Due to the experimental difficulties that we describe and bypass below, only a handful of measurements have been available for organic scintillators in the ``turnaround" 1-100 keV energy region, where the QF would be expected to display an abrupt monotonic increase with decreasing energy, in order to match the large values found using NE-110. Our dedicated measurement confirms that this is the case, validating the findings in \cite{ahlen1,ahlen2}, and the good prospects for use of hydrogenated targets in low-mass SIMP searches \cite{mypaper}. In combination with all previous data, our results also support an underlaying common physical basis for the production of light in all aromatic organic scintillators.

A second source of motivation for our work is the validation of QF models employed during neutron background studies performed within the COHERENT collaboration \cite{coherent}. We have employed EJ-301 to characterize the flux and energy distribution of prompt and neutrino-induced neutrons at the Spallation Neutron Source (SNS) of  the Oak Ridge National Laboratory. Unless controlled, these two sources of neutron backgrounds would be able to compete with Coherent Elastic Neutrino-Nucleus Scattering (CE$\nu$NS), a process recently measured for the first time at this site \cite{coherent2}. The measurements described below support the EJ-301 response model used in \cite{coherent2} to demonstrate a negligible neutron background contamination of the CE$\nu$NS signal.

\section{Neutron Source and Experimental Setup}

We follow a method previously implemented in a number of QF studies of  materials used in WIMP detectors. It consists of exposing a target detector under investigation to a collimated beam of monochromatic neutrons, while registering scattered neutrons with a backing detector (Fig.\ 1). Knowledge of initial neutron energy and scattering angle $\theta$ is sufficient to define the nuclear recoil (NR) energy deposited in the target, whenever this target is small enough to guarantee single scatters within. Delayed coincidences between target and backing detector, separated by neutron time-of-flight (TOF), are used to isolate these energy depositions. Gamma sources are typically used to establish an energy scale for electron recoils (ERs) in the target detector. In scintillators, a comparison of the light yield for ERs and NRs of same energy leads to the determination of the QF. A wide range of NR energies can be studied by varying neutron source energy and/or the angular position of the backing detector(s).

There are several difficulties involved in the use of this method for proton recoils below 100 keV. First, the neutron  energy necessary for this is modest, regardless of value of $\theta$. This precludes the use of organic scintillators with neutron/gamma pulse-shape discrimination (PSD) for the backing detector, an ideal choice able to reduce the severe gamma background contamination generated by most neutron sources. For the low neutron energies required, the majority of depositions in the backing detector would fall below the few tens of keV necessary for optimal PSD \cite{luo,ronchi}. In order to bypass this problem, we employed a large 5 cm diameter x 1.5 cm length $^{6}$LiI[Eu] scintillator, isotopically enriched to $>$96\% $^{6}$Li \cite{proteus}.  The  $^{6}$Li(n,$\alpha$) neutron absorption reaction produces a well-defined signal at 3.1 MeV electron-equivalent energy, distinguishable above the energy of most gamma backgrounds. We have successfully employed this alternative backing detector before, during studies of sub-keV nuclear recoils in germanium \cite{jcap}, using a 24 keV neutron beam \cite{kansas}.  Our present choice of neutron  energy at 245 keV was intentional, to profit from a broad resonance in the  $^{6}$Li(n,$\alpha$) cross-section  \cite{lii}. MCNP-PoliMi  simulations \cite{polimi} indicated that approximately 4\% of neutrons entering the front of our backing detector would produce the characteristic signal sought, which was deemed sufficient.

\begin{figure}[!htbp]
\includegraphics[width=0.46\textwidth]{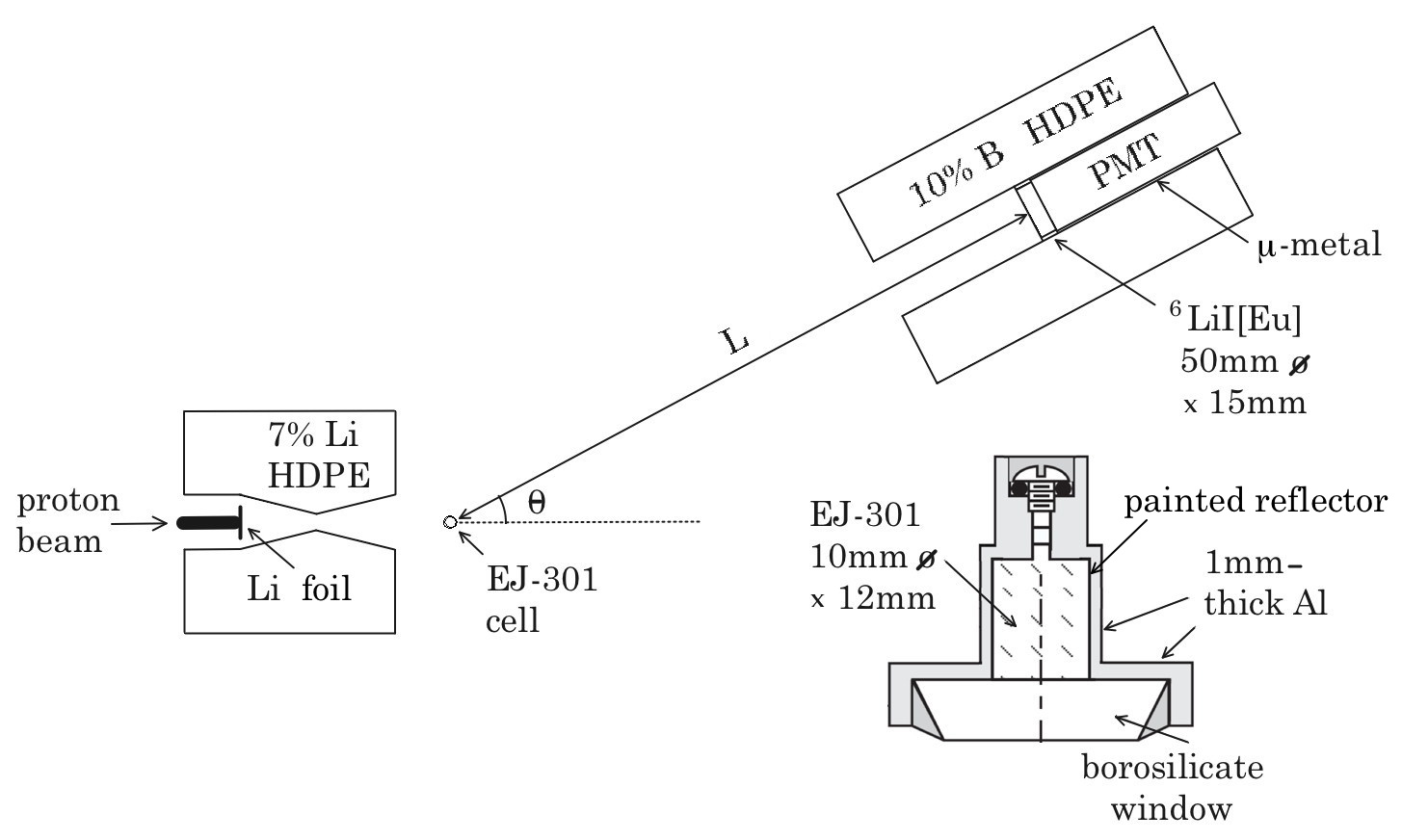}
\caption{\label{fig:lee} Experimental arrangement, with all dimensions to scale, derived from the MCNP-PoliMi simulation geometry. The scattering angle $\theta$ and distance $L$ between the $^{6}$LiI[Eu] backing detector and EJ-301 are indicated. A cross section of the small EJ-301 liquid scintillator cell is also shown.}
\end{figure}

\begin{widetext}
\begin{table}[h!]
\centering
\begin{tabular}{| c | c | c | c | c | c | c |} 
 \hline
 $\theta$($^{\circ}$) & recoil energy(keV) & L(cm) & exposure time(m) & integrated charge($\times10^{14}$C) & \# of events & quenching factor(\%)\\ [.7ex] 
 \hline
15.0$\pm$0.2 & 17.2$\substack{+3.8 \\ -5.5}$ & 60.2$\pm$0.4 & 140 & 0.84 & 275 & 12.30$\substack{+0.80 \\ -0.70}$\\[.7ex] 
19.7$\pm$0.4 & 28.1$\substack{+6.3 \\ -7.1}$ & 60.0$\pm$0.3 & 242  & 1.30 & 416 & 9.31$\substack{+0.45 \\ -0.40}$\\[.7ex] 
23.0$\pm$0.3 & 37.5$\substack{+5.9 \\ -7.0}$ & 59.5$\pm$0.4 & 180 & 0.98 & 306 & 10.41$\substack{+0.37 \\ -0.38}$\\[.7ex]
24.9$\pm$0.3 & 43.5$\substack{+8.4 \\ -8.5}$ & 60.1$\pm$0.5 & 210 & 1.23 & 343 & 8.65$\substack{+0.35 \\ -0.36}$\\[.7ex]
27.9$\pm$0.3 & 52.6$\substack{+9.3 \\ -8.1}$ & 60.0$\pm$0.5 & 115 & 0.74 & 152 & 8.95$\substack{+0.52 \\ -0.47}$\\[.7ex] 
30.2$\pm$0.3 & 60.0$\substack{+10.0 \\ -8.7}$ & 59.5$\pm$0.4 & 250 & 1.57 & 353 &  8.20$\substack{+0.30 \\ -0.25}$\\[.7ex]
33.1$\pm$0.5 & 69.7$\substack{+14.1 \\ -6.7}$ & 59.5$\pm$0.7 & 201 & 1.18 & 167 & 8.30$\substack{+0.42 \\ -0.36}$\\[.2ex] 
 \hline
\end{tabular}
\parbox{17cm}{\caption{Parameters corresponding to the seven angular measurements performed (see text). The ''number of events" column lists the integrated number of proton recoils with $<$25 PE in background-subtracted residual spectra (Fig.\ 4).}}
\label{table:1}
\end{table}
\end{widetext}

A second obstacle arises from the small light yield expected from protons in this low-energy region. This concern was addressed through use of a Hamamatsu H11934-200 Ultra-Bialkali (UBA) photomultiplier (PMT)  to monitor the emissions from the EJ-301 target cell. At the 425 nm peak emission wavelength of EJ-301, the UBA photocathode exhibits a 54\% higher quantum efficiency than a standard bialkali PMT \cite{uba}. The PMT was directly coupled using optical RTV to a small ($\sim$1 c.c.) custom-built EJ-301 cell (Fig.\ 1). MCNP-PoliMi  simulations indicate that only 14\% of interactions in this small cell lead to more than a single scatter per incoming neutron.  These few multiple scatters are taken into account in the analysis below. Excessive multiple scatters in a larger detector can lead to systematics able to affect QF determination.

The measurement took place at the Tandem Van De Graff accelerator of the Triangle Universities Nuclear Laboratory \cite{tandem}. This facility uses a direct extraction negative ion source to produce H$^{-}$ and D$^{-}$, which are then accelerated and converted into their positively charged counterparts inside of the tandem \cite{tandem2}. The accelerator is capable of energies up to 20 MeV, operating in either direct-current (DC) or pulsed mode. A series of bending magnets constrain the energy of the beam and direct it to a suitable neutron-generating material, in our case 650 nm of lithium fluoride (LiF) evaporated onto a 125 $\mu$m-thick  tantalum foil (Fig.\ 1), a substrate chosen to minimize the production of gamma rays. Our experiment ran in DC mode and delivered approximately one microamp of protons-on-target (POT) over its duration. The total charge delivered to the foil during each run was recorded at the accelerator console (Table I).  A proton energy of 2 MeV produces monochromatic $\sim$245 keV neutrons in the forward direction, via the  $^{7}$Li(p,n) reaction \cite{tandem3,tandem4}. A precise neutron energy characterization was determined by switching the proton beam to pulsed mode, and measuring the difference in TOF between gammas and neutrons, traveling from the lithiated foil to a plastic scintillator (Fig.\ 2).

\begin{figure}[!htbp]
\includegraphics[width=0.43\textwidth]{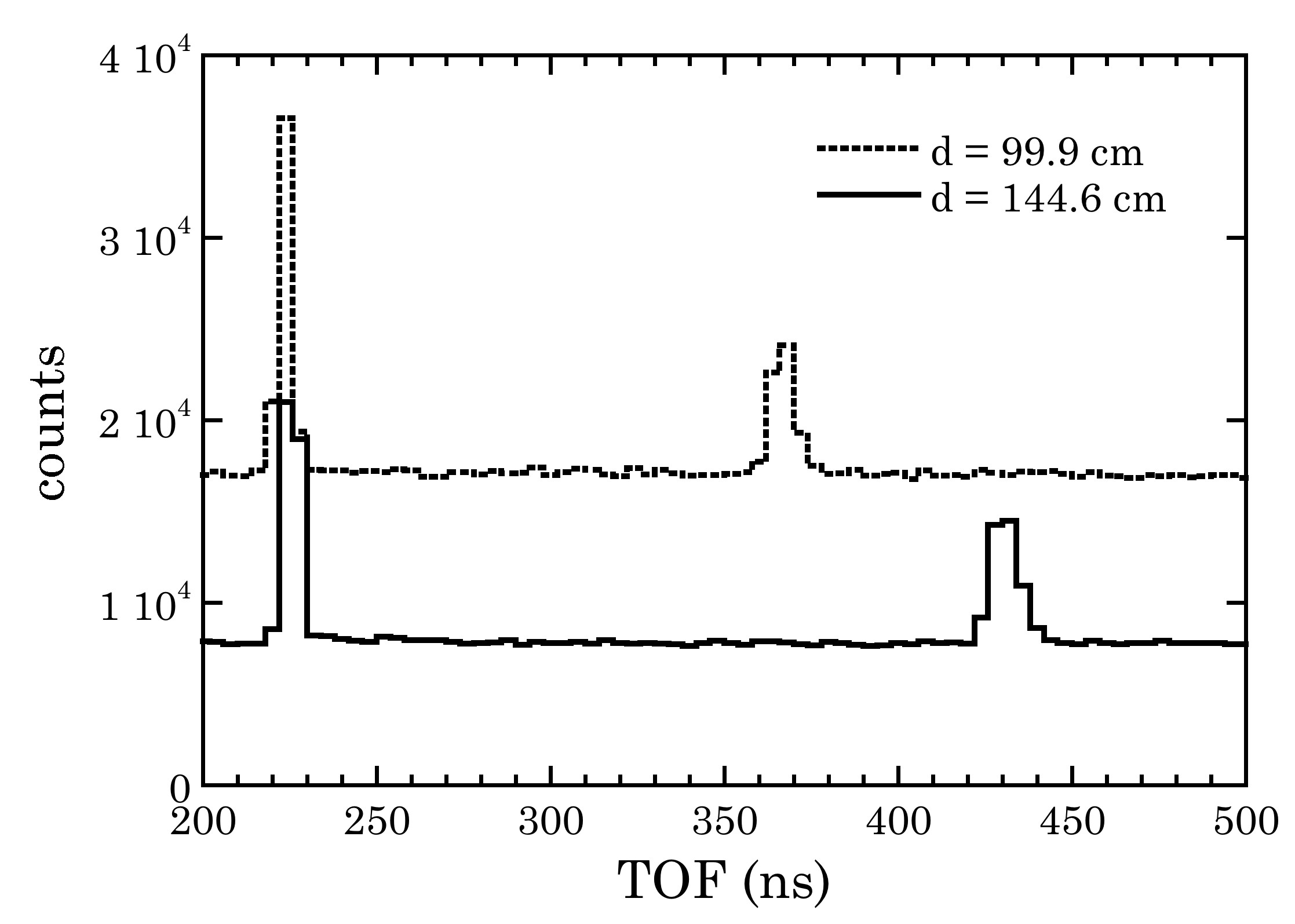}
\caption{\label{fig:lee} TOF for gammas (left peaks) and neutrons (right peaks) produced in the lithium foil, arriving to plastic scintillator placed in the forward ($\theta$ = 0$^{\circ}$) direction. TOF is relative to a logic signal in phase with POT pulses. Two distances $d$ between foil and scintillator were tested. For each, neutron velocity $v$ can be extracted from the expression $d/v - d/c = \Delta t$, where $c$ is the speed of light and  $\Delta t$ is the n-$\gamma$ difference in TOF.  Both measurements agree within 1\%, yielding an average neutron kinetic energy of 244.6 $\pm$ 8.4 keV. }
\end{figure}

A high-density polyethylene (HDPE) collimator surrounding the neutron source was 7\% enriched in natural lithium, to reduce the capture gamma background reaching the EJ-301.  The target cell was placed in close proximity (19.5 cm) to the source, to maximize the neutron flux at its site. A double-conical tapering of the collimator (Fig.\ 1) blocks the line-of-sight between neutron source and backing detector, while minimizing the flux of  neutrons moderated in the collimator that can reach the EJ-301 cell. A borated cylindrical shield around the backing detector served the purpose of reducing spurious backgrounds from stray thermal neutron capture in $^{6}$LiI[Eu]. The gain of the PMT monitoring the backing detector was stabilized against a $\theta$ dependence through the addition of $\mu$-metal magnetic shielding. Laser tools were used for component alignment. Utilizing plumb bobs, projections on a two-dimensional grid drawn on the ground were used to estimate the small uncertainties in $\theta$ and $L$ (Table I).

\section{Data Acquisition and Analysis}

One of two identical signals from the base of the backing detector PMT was amplified, shaped, and routed through a single-channel analyzer set to generate an output  for pulses in the 2.1-4.1 MeV energy interval, i.e., centered around the $^{6}$LiI[Eu] (n,$\alpha$) signal. This output is used as an external trigger to an Acqiris DP1400 fast digitizer. It recorded the second $^{6}$LiI[Eu] output and EJ-301 PMT signals, at a sampling rate of 500 MS/s, over sufficiently long traces preceding the trigger. A -900V bias was applied to the EJ-301 PMT, high enough to provide SPE sensitivity. Traces were bundled and stored to disk. The triggering rate was well-below the maximum throughput of this system, for all values of $\theta$. 

\begin{figure}[!htbp]
\includegraphics[width=0.46\textwidth]{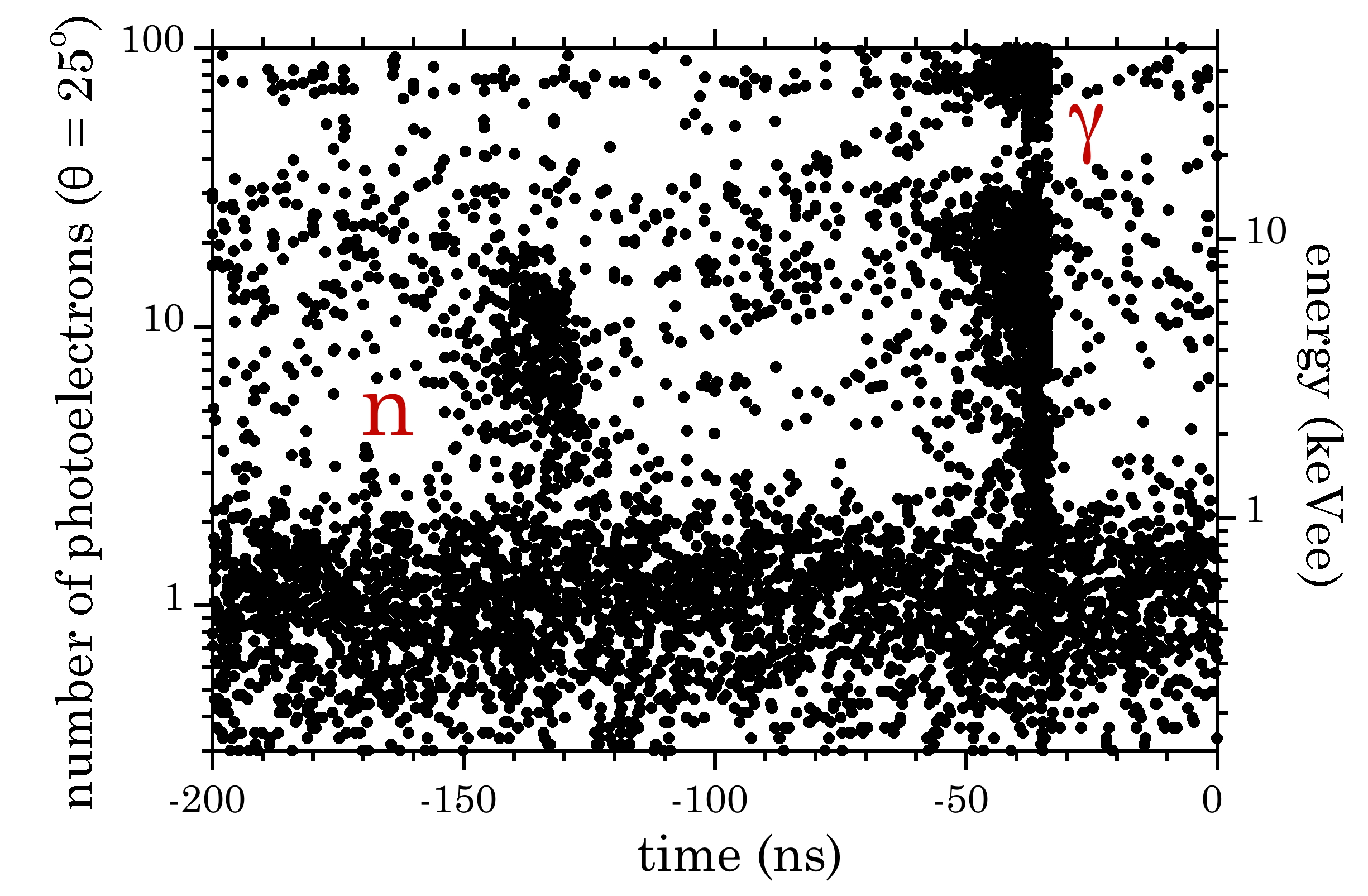}
\caption{\label{fig:lee} Signals in EJ-301, time-referenced to the onset of signals in the backing detector. The horizontal coordinate shows the onset of EJ-301 light emission, and the vertical its light yield integrated over 80 ns. As in Fig.\ 2, a time difference of $\sim$90 ns between the indicated n and $\gamma$ populations matches the expected 245 keV neutron TOF over the distance $L$ = 60 cm between detectors. The $\gamma$ offset with respect to t = 0 is due to a rise-time to reach an analysis threshold for  $^{6}$LiI[Eu] signals. Variations by $\sim$20 ns in the position of n signals were observed for different values of $\theta$, as expected from neutron  energy losses in EJ-301 (Table I). }
\end{figure}

An ER energy scale for EJ-301 was established using the 59.5 keV gamma emission from an $^{241}$Am source, integrated over a 80 ns window. A light yield of 2.1 PE/keV was obtained. A small non-linearity of order 5\% in the response of EJ-301 to 10-100 keV ERs is neglected in our analysis (EJ-301 is identical in formulation to the BC-501A studied in this respect in \cite{nonlin}). 

\begin{figure}[!htbp]
\includegraphics[width=0.45\textwidth]{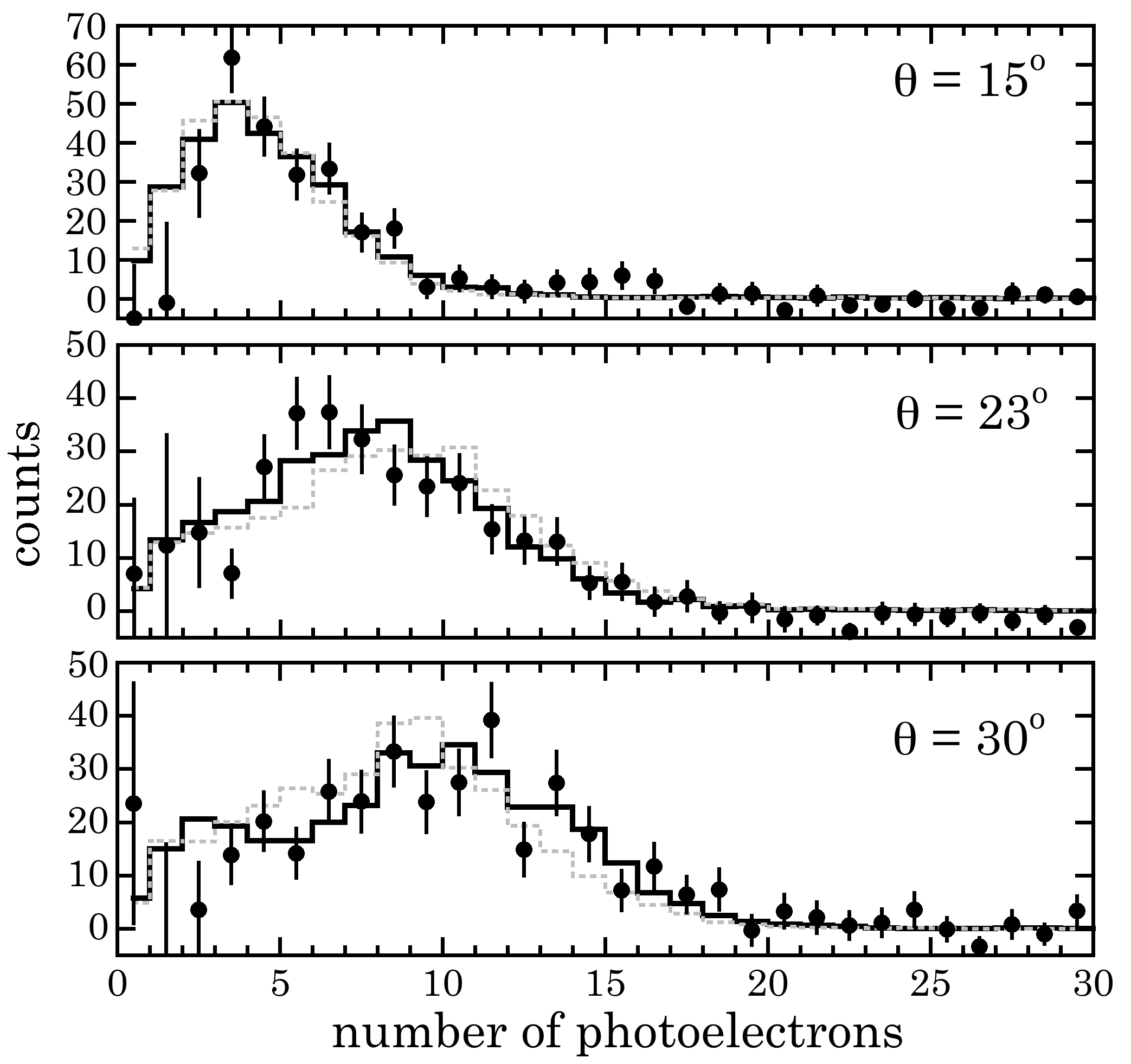}
\caption{\label{fig:lee} Examples of background-subtracted residual spectra from proton recoils in EJ-301. The  electron-equivalent energy scale, defined via $^{241}$Am gamma calibration, is 2.1 PE/keV. Solid histograms correspond to simulated spectra for best-fit QF values of 12.3\%, 10.4\%, 8.2\%, top to bottom. The sensitivity of these fits is illustrated by dashed greyed histograms for QF = 11.3\%, 11.4\%, 7.2\%, respectively. }
\end{figure}

An offline analysis code was used to extract the amplitude of EJ-301 pulses preceding the onset of $^{6}$LiI[Eu] trigger signals. Prompt coincidences from gamma scattering affecting both detectors were observed, as well as delayed coincidences from neutron interactions. The latter resided in the [-160,-110] ns interval of Fig.\ 3, for all values of $\theta$ tested. A second interval [-660,-160] ns was used to characterize a time-independent background of spurious coincidences, consisting mainly of SPEs. Following normalization to the same time span, the residual spectrum of signals falling in these two time intervals provides a background-free picture of energy depositions from proton recoils in EJ-301 (Fig.\ 4).

A detailed MCNP-PoliMi geometry of the setup and detectors was constructed. Neutrons from a 244.6$\pm$8.4 keV point source at the lithiated foil were used as simulation input, with emissions restricted to a $\pm 5^{\circ}$ forward cone, uniformly sampled. This maximizes the efficiency of the simulation, while correctly approximating the neutron angular distribution \cite{tandem4}, and including the effect of the collimator.  Post-processing of the output extracts proton recoil energies in EJ-301 for neutron histories producing (n,$\alpha$) reactions in the backing detector.  For each value of $\theta$, a fine-grained sampling of QF values within the interval 2-20\% were tested by the simulation, as follows. Individual proton recoil energies were converted into an expected PE yield through a choice of QF, and the electron-equivalent yield of 2.1 PE/keV. Poisson fluctuations around this expectation value generate a simulated number of PE for each NR. Multiple scatters and infrequent carbon recoils (occurring for 16\% of   histories) were included. However, the smaller maximum recoil energy of carbon nuclei and a modest $\sim$1\% quenching factor  \cite{spooner,ej301quenching} renders their contribution negligible. 

\begin{figure}[!htbp]
\includegraphics[width=0.4\textwidth]{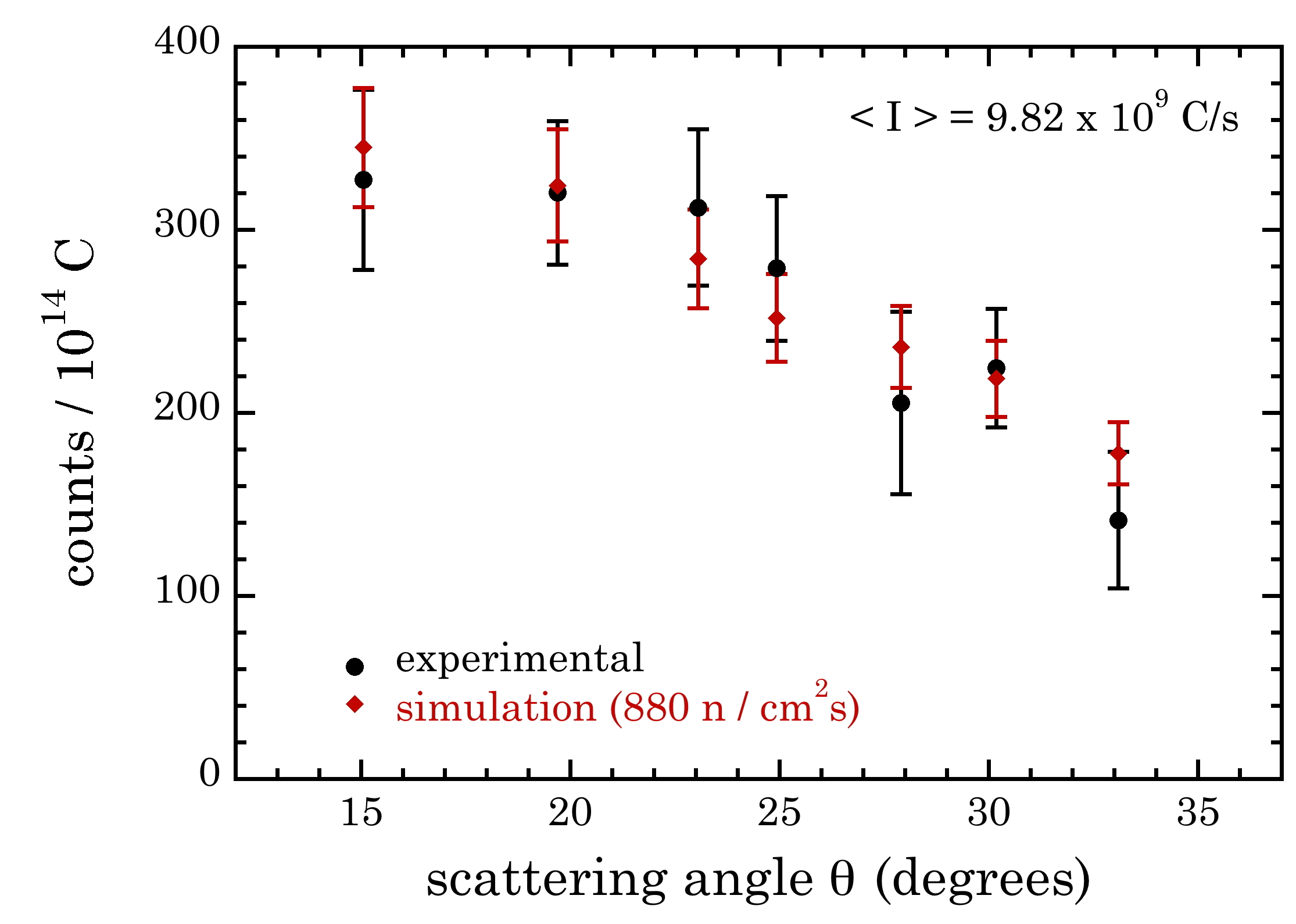}
\caption{\label{fig:lee} Comparison of experimental and simulated proton recoil rates in EJ-301. Experimental rates are normalized to the indicated reference proton current delivered to the lithiated foil, using charge, number of events, and exposure times in Table I. Simulated rates are for the  best-fit neutron flux at the position of the EJ-301, stated in the label.  }
\end{figure}

Simulated residual spectra were normalized to the same number of events yielding $<$ 25 PE as in the experimental residuals (Fig.\ 4, Table I). A profile likelihood estimator returned the QF providing the best fit between experimental and simulated residuals, and its asymmetric one-sigma confidence interval (Table I, vertical errors in Fig.\ 6).  For each $\theta$, the span of proton recoil energies probed by the measurement was extracted from the asymmetric half-width at half-maximum of the simulated distribution of recoil energies. This is listed as an energy uncertainty in Table I, and shown as horizontal error bars in Fig.\ 6. For this analysis, the QF was treated as approximately constant over each of these energy spans.

The simulation was also used to find the neutron flux at the EJ-301 position that best matched the experimentally observed proton-recoil rates. Following the normalization of these rates to a reference current delivered to the lithiated foil, an excellent agreement between simulation and measurement was noticed for all values of $\theta$ (Fig.\ 5), when a best-fitting neutron flux of 880 n/cm$^{2}$s is adopted. This flux matched estimates based on previous operation of this neutron source. The agreement  visible in Fig.\ 5 provides an important cross-check on the absence of systematics affecting our QF determination at the lowest recoil energies probed. Neglecting this control comparison can lead to overestimated low-energy QF values \cite{systematics}.

\begin{figure}[!htbp]
\includegraphics[width=0.45\textwidth]{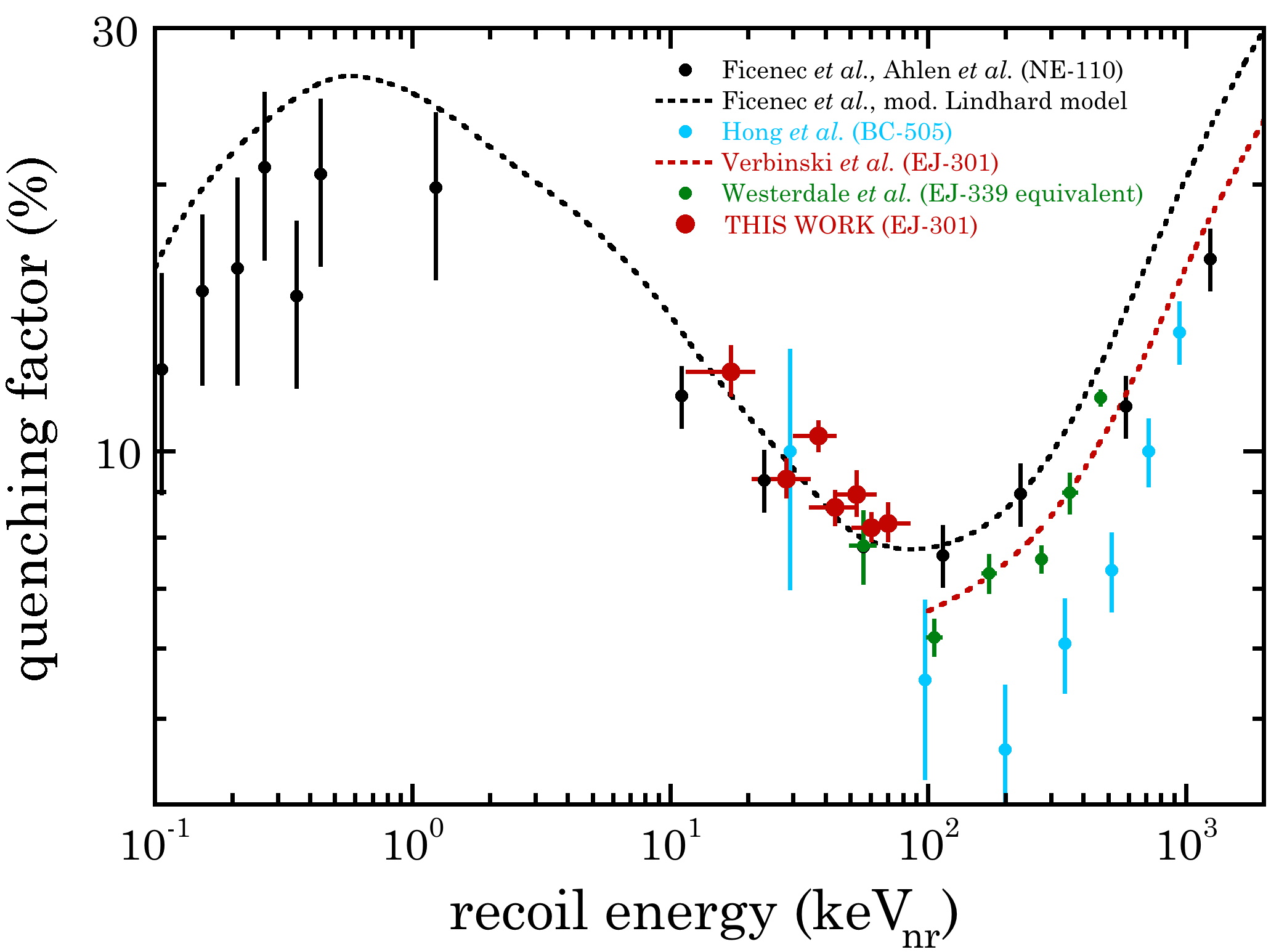}
\caption{\label{fig:lee} Low-energy quenching factors for proton recoils in organic scintillators \protect\cite{ahlen1,ahlen2,spooner,ej301quenching,calaprice}, including present results. A dotted black line represents the modified Lindhard model proposed in \protect\cite{ahlen2}, and adopted by the COHERENT collaboration for neutron background studies in \protect\cite{coherent2}. Proton recoil light yields from \protect\cite{ahlen2} are converted here to a quenching factor via a 3 eV mean photon energy for NE-110 scintillation \protect\cite{ahlen1}, and a 9.2 photon/keV scintillation light yield for ERs in this material \protect\cite{ej208}.}
\end{figure}

\section{Results and Conclusions}

Our new QF measurements are shown in Table I and Fig.\ 6. The figure also displays all previously available data for low-energy proton recoils in aromatic organic scintillators. As mentioned in Sec.\ I, our present measurements confirm the EJ-301 response used to demonstrate a near-complete absence of  neutron backgrounds in a first CE$\nu$NS measurement \cite{coherent2}. The ascending trend in QF found below 100 keV supports the measurements below 1 keV described in \cite{ahlen1,ahlen2}, and bodes well for planned use of hydrogenated scintillators in low-mass dark matter searches \cite{mypaper}.

The similar behavior noticeable in Fig.\ 6 for all aromatic organic scintillator formulations suggests a common physical basis in their production of light (namely, the excitation and transitions of $\pi$-electronic states in benzenic rings \cite{brooks}). Specifically, and similarly to what was recently reported in \cite{calaprice}, our data are much better described by a modified Birks' model of scintillation production containing a quadratic dependence on proton total stopping power \cite{birksmod1,birksmod2}, than by a standard Birks' model \cite{birks}. This preference for the modified model can be quantified at the 5.4 sigma level, under a standard likelihood ratio test.

\begin{acknowledgments}
We are indebted to Chris Maxwell and Chuck Hurlbut at Eljen Technology for their development of the miniature EJ-301 cell used in this study, and to John Dunham for his assistance setting up the neutron source. This work was supported by NSF award PHY-1506357.
\end{acknowledgments}

\end{document}